\documentclass[intlimits,twoside,a4paper]{article}

\usepackage[cp1251]{inputenc}

\usepackage[eqsecnum]{cmpj3}
\usepackage[mathscr]{eucal}

\usepackage{bm}

\issue{2020}{23}{3}{33708}
\doinumber{10.5488/CMP.23.33708}
\title[Interaction of electrons with acoustic phonons in AlN/GaN resonant tunnelling nanostructures ]%
{Interaction of electrons with acoustic phonons in AlN/GaN resonant tunnelling nanostructures at different temperatures%
}%

\author[I.V. Boyko, M.R. Petryk]{I.V. Boyko, M.R. Petryk}
\address{Ternopil Ivan Puluj National Technical University, 56 Ruska St., 46001 Ternopil, Ukraine}
\date{Received June 12, 2020, in final form July 21, 2020}

\begin{document}

\maketitle

\begin{abstract}

The theory of the interaction of electrons with acoustic phonons in multilayer nitride-based AlN/GaN nanostructures was developed for the first time at $T\geqslant 0$ using the method of finite-temperature Green's functions and Dyson equation. Components of the Hamiltonian describing the system of electrons with acoustic phonons and the magnitudes of the electron spectrum shifts due to the electron-phonon interaction were obtained. Dependences of the electronic spectrum levels and spectrum of the acoustic phonons were found depending on the position of the internal potential barrier in the studied nanostructure. The temperature shifts of the electronic spectrum and decay rates were calculated for various values of temperature $T$.
\keywords acoustic phonon, electron-phonon interaction, Green's function, Dyson equation, nitride-based nanostructure
%
\end{abstract}

\section{Introduction}

In modern nanotechnology and in related areas of nanoscience, special attention is paid to the study of multilayer resonant tunnelling structures (RTS) created on the basis of binary ($\rm GaN$, $\rm AlN$) and ternary ($\rm GaAlN$) alloys of nitride-based semiconductor materials. The mentioned nanostructures are widely used as active elements of cascades of quantum cascade lasers (QCL) \cite{Fujikawa2019,Li2019} and detectors (QCD) \cite{Lim2017,Mensz2019} operating in the near and middle ranges of infrared waves.

Lately, considerable attention of researchers of multilayer nitride nanostructures has been paid to the study of internal electric fields arising in RTS layers due to significant values of spontaneous and piezoelectric polarizations \cite{Bernardini1998,Bernardini1999}, and due to the development of methods for calculating potential profiles of these nanosystems \cite{Saha2016,Boyko2018}. Besides, some theoretical and experimental papers deal with the study of excitons and interband transitions in the mentioned nanosystems \cite{Bayerl2019,Staszczak2020}. Despite the fact that optical phonons and the interaction of electrons with them in nitride-based nanostructures of both multilayer plane nanosystems \cite{Yan2003}, quantum dots \cite{Yamanaka2008} and quantum wires \cite{Zhang2006} are studied well enough, but for the acoustic phonons such studies, in fact, are not available. An exception is a group of papers on the spectral characteristics of acoustic phonons performed by Pokatilov et al. \cite{Pokatilov2003,Pokatilov2004} for single-well nanosystems in a simplified model in which the components of the stress tensor ($\sigma_{ij},\, i,j=1,2,3$) were assumed to be zero at the boundaries with an external semiconductor medium or sapphire substrate. That is why it is not possible to apply this theory to multilayer RTS used as effective elements of the QCL and QCD cascades due to the requirements of consistency of their cascades.

Two similar papers \cite{Pokatilov2004,Yang2013} should be specified, where the interaction of electrons with acoustic phonons was studied based on the  one-well nanostructure models mentioned above \cite{Pokatilov2003,Pokatilov2004}. Besides, based on these papers, for single-well nanostructures, recently there been studied the influence of heterogeneity and prestress field effects, prestress and surface/interface stress, the piezoelectric effect on their thermal conductivity and phonon properties \cite{Zhu2016,Wang2018}.

The theory of acoustic phonons modes arising in multilayer nitride resonant tunnelling structures and their calculations were performed recently for the first time in papers \cite{Boyko2020} and \cite{Boyko2020-1} for binary and ternary semiconductor alloys, respectively. The results obtained in these papers demonstrate a significant difference in the calculated dependences of the spectra of acoustic phonons  and corresponding components of the displacement field which deal with the formation of an additional group of the spectrum of acoustic phonons  dependences on the wave vector in the presence of a layer with a ternary $\rm AlGaN$ semiconductor being in the nanostructure. Besides, for nanostructures created on the basis of binary semiconductors, the moduli of the maximum values of
 displacement field components   decrease with an increase of the level number of the spectrum of acoustic phonons, which is not observed when the $\rm AlGaN$ semiconductor layer is available.

Thus, it should be concluded that the theory of the interaction of electrons with acoustic phonons in multilayer nitride semiconductor RTS is not available nowadays.

In the presented paper, a theory of the interaction of electrons with acoustic phonons in a multilayer nitride RTS is developed. In the representation of the second quantization, the partial components of Hamiltonian which describes the interaction of electrons with various types of acoustic phonons at $T\geqslant0$ are obtained. Based on the analysis of the Green's functions poles, the temperature shifts of the electronic spectrum levels in the studied nanostructure and their dependence on its geometric parameters for different values of temperature $T$ are obtained.

\smallskip

\section{Electronic spectrum, wave functions and potential profile of the multilayer nitride-based nanostructure}

\begin{figure}[!b]
\centerline{\includegraphics[width=0.50\textwidth]{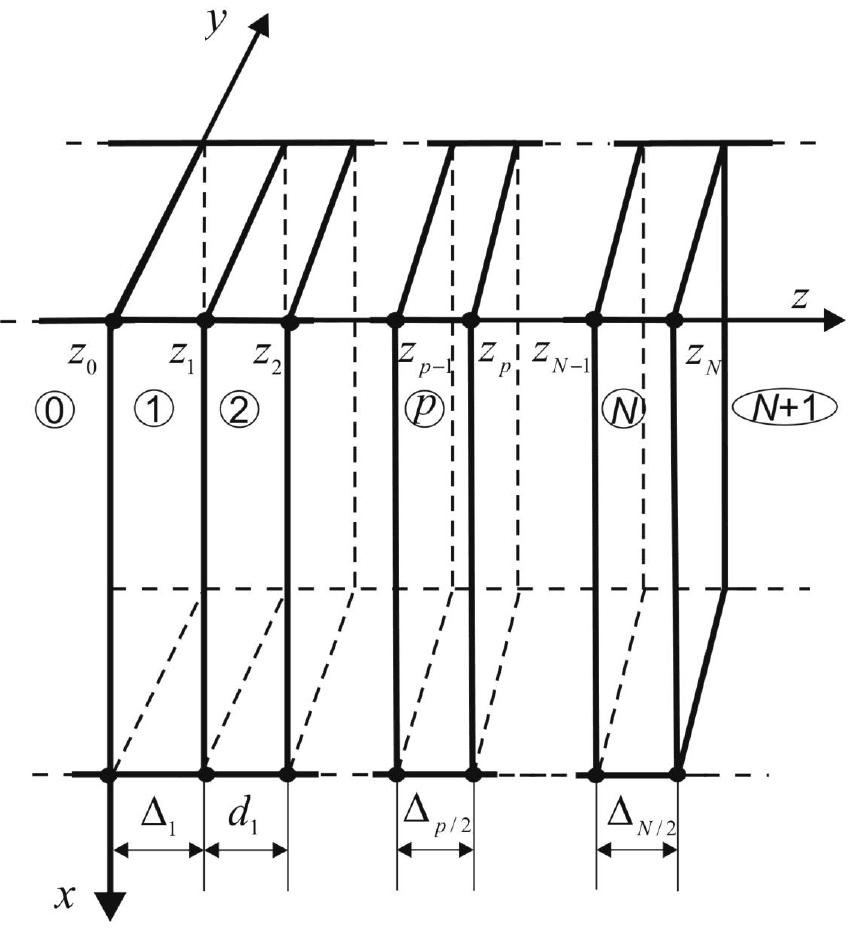}}
\caption{Geometrical scheme of the plane multilayer AlN/GaN nanostructure.} \label{fig-smp1}
\end{figure}

We study stationary electronic states in a plane multilayer semiconductor AlN/GaN nanostructure which works as a separately selected QCD cascade. The Cartesian coordinate system is chosen in such a way that its axis $Oz$ is perpendicular to the heterointerfaces between the media of the given $N$ layers of the nanosystem (figure~\ref{fig-smp1}). To ensure the consistency of the QCD cascades and the random selection of a separate cascade \cite{Boyko2020-1}, it is assumed that the media (0) and $(N+1)$, respectively, to the left and to the right of the nanosystem, correspond to the $ \rm AlN $ semiconductor medium, and the inner layers of the nanostructure are formed by alternating semiconductors $ \rm GaN$ and $\rm AlN$.

Using the effective mass model and the dielectric continuum model, the effective electron mass $m(z)$ and the dielectric permittivity $\varepsilon(z)$ of the RTS can be written as follows:
\begin{align}
&\begin{array}{l} {\left(\begin{array}{l} {m(z)} \\ \displaystyle {\varepsilon (z) } \end{array}\right)=\left(\begin{array}{l} {m^{(0)} } \\ {\varepsilon ^{(0)} } \end{array}\right)\theta (-z)+\left(\begin{array}{l} {m^{(N+1)} } \\ {\varepsilon ^{(N+1)} } \end{array}\right)\theta (z-z_{N} )+\displaystyle \sum _{p=1}^{N}\left(\begin{array}{l} {m^{(p)} } \\ \displaystyle {\varepsilon ^{(p)} } \end{array}\right)\left[\theta (z-z_{p-1} )-\theta (z-z_{p} )\right],} \nonumber\end{array}\\
& m^{(0)} =m^{(N+1)} =m_{(1)} ,\, \, \varepsilon ^{(0)} =\varepsilon ^{(N+1)} =\varepsilon _{(1)} ,\, \, m^{(p)} =\left\{\begin{array}{l} {m_{(0)} ,} \\ {m_{(1)} ,} \end{array}\right. \, \, \varepsilon ^{(p)} =\left\{\begin{array}{l} {\varepsilon _{(0)} ,\, \, p-\text{even}} \\ {\varepsilon _{(1)} ,\, \, p-\text{odd}} \label{eq2.1}\end{array},\right.
\end{align}
where $\theta (z)$   is the Heaviside step function,  $m_{(0)} ,\,m_{(1)}$  are values of effective electron masses in potential barriers and wells, respectively, $\varepsilon _{(0)} ,\, \, \varepsilon _{(1)}$  are  dielectric permittivities of  semiconductor layers of the nanostructure, correspondingly. Designations with upper indices ($m^{(p)},\,\varepsilon^{(p)}$) are used to correctly represent the values of the effective masses and the  dielectric permittivities of an arbitrary layer of the nanostructure. Their definition is visible at the end of the formula (\ref{eq2.1}).

Considering the fact that the geometric dimensions of the RTS cross-section by plane $xOy$ exceed the longitudinal dimensions of the nanostructure ($l_{x},l_{y}\gg z_{N}$), it is advisable to represent the wave function of the electron in the form similar to the Bloch function:
\begin{align}
\Psi _{E\, \bar{k}_{0} } (\bar{r}_{0} ,z)=\frac{1}{\sqrt{l_{x} l_{y} } } \re^{\ri(\bar{k}\bar{r}_{0}) } \Psi _{E} (z),
\label{eq2.2}
\end{align}
where $\bar{r}_{0}$ and $\bar{k}$ are the vector in $xOy$ plane and the quasimomentum of the electron, respectively.

Separation of the motion of an electron in the direction along the $Oz$ axis from its motion in the plane
$xOy$ is performed taking into account relation (\ref{eq2.2}), as it was fulfilled, for example,  in papers \cite{Tkach2014,Gao2007}. The stationary spectrum of the electron $E_{n}$  and its wave functions $\Psi_{E} (z)$ are now obtained by finding solutions of a self-consistent system of Schr\"odinger-Poisson equations:
\begin{align}
\left\{\begin{array}{l} \displaystyle {-\frac{\hbar ^{2} }{2} \frac{\rd}{\rd z} \left[\frac{1}{m(z)} \frac{\rd\Psi (z)}{\rd z} \right]+V(z)\Psi_{E} (z)=E\Psi_{E} (z),} \\ \\ \displaystyle{\frac{\rd}{\rd z} \left[\varepsilon (z)\frac{\rd V_{H} (z)}{\rd z} \right]=-e\rho ^{(\text{elect})} (z)}, \end{array}\right.
\label{eq2.3}
\end{align}
where the nanostructure effective potential for the electron is the sum of such partial components \cite{Saha2016,Boyko2018}:
\begin{align}
V(z)=\Delta E_{C} (z)+V_{E} (z)+V_{H} (z)+V_{HL} (z),
\label{eq2.4}
\end{align}
the analytical form and the meaning of which will be established further.
The total electron energy in the RTS is defined as follows:
\begin{align}
E_{n\bar{k}} =E_{n} +\frac{\hbar ^{2} k^{2} }{2m_{n}^{{\rm (eff)}} },
\end{align}
where the second term describes the energy, corresponds to the movement of the electron in the direction perpendicular to the $Oz$  axis (in the $xOy$ plane), and the correlated in-plane effective mass of the electron $m_{n}^{{\rm (eff)}}$  for $\Gamma$-conduction subband $n$ is obtained taking into account (\ref{eq2.1}). It provides the approximation of the contribution of all the nanostructure layers \cite{Tkach2014,Gao2007}:
\begin{align}
m_{n}^{{\rm (eff)}} =\left[\int _{-\infty }^{+\infty }\left(\left|\Psi _{n} (z)\right|^{2} /m(z)\right)\rd z \right]^{-1}.
\end{align}
The density of charges in the RTS is defined as follows:
\begin{align}
\rho ^{(\text{elect})} (z)=e[N_{D}^{+} -n(z)]+\sum _{p=1}^{N}\sigma _{p} \delta (z-z_{p} )\,,
\end{align}
where $N_{D}^{+}$ is concentration of ionized donor impurities,
\begin{align}
n(z)=\frac{m(z)k_{\text{B}} T}{\piup \hbar^{2} } \sum _{n}\left|\Psi (E_{n} ,\, \, z)\right|^{2}  \ln \left|1+\exp \left[(E_{F} -E_{n} )(k_{\text{B}} T)^{-1} \right]\right|
\label{eq2.5}
\end{align}
is the concentration of electrons determining the static space charge in the RTS,  $E_{F}$  is the Fermi level of nanosystem material, $E_{n}$  are the stationary electronic spectrum energy values, $\sigma _{p} =P_{p+1} -P_{p}$   is the surface density of charges arising due to different polarizations in adjacent layers of the RTS, $\delta(z)$  is the Dirac delta function. Here, $P=P_{Sp} +P_{Pz}$ is total value of polarization arising in the layers of the RTS, $P_{Sp} ,\, \, P_{Pz}$  are spontaneous and piezoelectric polarization, respectively.
In the expression (\ref{eq2.4}):
\begin{align}
\displaystyle \Delta E_{C} (z)=\left\{\begin{array}{l} {0,\, \, \text{wells}} \\ \\\displaystyle {0.765[E_{g} ({\rm AlN})-E_{g} ({\rm GaN})],\, \, \, \, \text{barriers}} \end{array}\right.
\label{eq2.9}
\end{align}
 is the potential profile of a nanosystem for an electron calculated without taking into account the internal electric fields. The dependence of the band gap $E_{g}$ on temperature $T$ in relation (\ref{eq2.9}) can be calculated using the linear-quadratic Varshni relation \cite{Saha2016,Boyko2018,Piprek2007} for a semiconductor alloy:
\begin{align}
E_{g} (T)=E_{g} (0)-\frac{aT^{2} }{b+T}\,,
\end{align}
where  $E_{g} (0)=E_{g}^{AlN} (0)$  is the band gap for a AlN semiconductor at $T=0$~K, $a={\rm 1.799}$~meV/K, $b=1462$~K  are the Varshni parameters \cite{Piprek2007}.

The effective potential component $V_{E}(z)$  is determined by interaction of the electron with internal electric fields ($F_{p} ,\, \, p=1\ldots N)$ arising in the RTS due to the existence of spontaneous and piezoelectric polarizations in its layers \cite{Bernardini1998,Bernardini1999,Boyko2018}:
\begin{align}
 V_{E} (z)&= e\sum _{p=1}^{N}(-1)^{p-1} (F_{p} z-F_{p-1} z_{p-1} )\left[\theta (z-z_{p-1} )-\theta (z-z_{p} )\right] ,\nonumber\\ 
 F_{0} &=0;\quad F_{p} =\sum _{k=1,\, k\ne p}^{N}\left(P_{k} -P_{p} \right)(z_{k} -z_{k+1} )/\varepsilon ^{(k)}  \left/{\vphantom{\sum _{k=1,\, k\ne p}^{N}\left(P_{k} -P_{p} \right)(z_{k} -z_{k+1} )/\varepsilon ^{(k)}   \varepsilon ^{(p)} \sum _{k=1}^{N}(z_{k} -z_{k+1} )/\varepsilon ^{(k)}  }}\right. \varepsilon ^{(p)} \sum _{k=1}^{N}(z_{k} -z_{k+1} )/\varepsilon ^{(k)}.\label{eq2.11} 
\end{align}
The effective potential component:
\begin{align}
\begin{array}{l} \displaystyle {V_{\text{HL}} (z)=-\frac{1}{4\piup } \left(\frac{9}{4\piup ^{2} } \right)^{{1 \mathord{\left/{\vphantom{1 3}}\right.\kern-\nulldelimiterspace} 3} } \left[1+\frac{0,6213r_{s} }{21} \ln \left(1+\frac{21}{r_{s} (z)} \right)\right]\frac{e^{2} }{\varepsilon _{0} r_{s} (z)\varepsilon (z)a_{\text{B}}^{*} (z)} \,, } \\ \\ \displaystyle {r_{s} (z)=\left(4\piup a_{\text{B}}^{*3} n(z)/3\right)^{-{1 \mathord{\left/{\vphantom{1 3}}\right.\kern-\nulldelimiterspace} 3} } ,\quad a_{\text{B}}^{*} (z)=\varepsilon (z)/m(z)a_\text{B} } \end{array}
\end{align}
is the Hedin-Lundquist exchange-correlation potential \cite{Hedin1971}, where $a_\text{B}$  is the Bohr radius.

The solution of the Schr\"odinger-Poisson system of equations in calculating the potential profiles of nanosystems was determined mainly numerically \cite{Saha2016}. Then, using the approach used in the paper \cite{Boyko2018}, the potential $V_{H} (z)$ determined by the contribution of the charge carriers within the RTS, can be represented in an analytical form:
\begin{align}
V_{H} (z)&= \sum _{p=1}^{N}V_{H}^{(p)} (z) \left[\theta (z-z_{p-1} )-\theta (z-z_{p} )\right];\quad V_{H}^{(p)} (z)=-\frac{e}{\varepsilon _{0} \varepsilon ^{(p)} } \int _{0}^{z}\int _{0}^{\xi _{2} }\left\{e\left(N_{D}^{+} -\frac{m^{(p)} k_{\text{B}} T}{\piup \hbar ^{2} } \right.  \right.\nonumber\\
&\left. \times \left.  \sum _{n}\left|\Psi (E_{n} ,\, \, \xi _{1} )\right|^{2}  \ln \left|1+\exp \left(\frac{E_{F} -E_{n} }{k_{\text{B}} T} \right)\right|\right)+\sigma _{p} \delta (\xi _{1} -z_{p} )\right\}\rd \xi _{1} \rd \xi _{2}\,,
\end{align}
where the integral is determined exactly.

Now, the solution of system (\ref{eq2.3}) can be found using the iteration method \cite{Boyko2018}, successively finding solutions of the Schr\"odinger equation and each time approximating the value of the found effective potential (\ref{eq2.4}) by $U_{{\rm appr}} (z)=V(z)$, where:

\begin{eqnarray}
&&U_{{\rm appr}} (z)=\sum _{p=1}^{N}\sum _{l=0}^{M}\left[(V(z_{p_{l+1} } )-V(z_{p_{l} } ))/(z_{p_{l+1} } -z_{p_{l} } )\right]  z\left[\theta (z-z_{p_{l} } )-\theta (z-z_{p_{l+1} } )\right],\label{eq2.15}\\
&&\Psi (E,z)=A^{(0)} \re^{\chi ^{(0)} (z)z} \theta (-z)+\sum _{p=1}^{N}\sum _{l=0}^{M}\left[A^{(p_{l} )} Ai(\zeta ^{(p_{l} )} (z))+B^{(p_{l} )} Bi(\zeta ^{(p_{l} )} (z))\right] \nonumber\\
&&\times\left[\theta (z-z_{p_{l} } )-\theta (z-z_{p_{l+1} } )\right] + 
B^{(N+1)} \re^{-\chi ^{(N+1)} (z)z} \theta (z-z_{5} ),\, \, \zeta ^{(p_{l} )} (z)\nonumber\\
&&=\left[2m^{(p_{l} )} eF(z_{p_{l} } )/\hbar ^{2} \right]^{{1 \mathord{\left/{\vphantom{1 3}}\right.\kern-\nulldelimiterspace} 3} } \left[(\Delta E_{C} (z)-E)/eF(z_{p_{l} } )-z\right],\nonumber\\ 
&&\chi ^{(0)} (z)=\chi ^{(N+1)} (z)=[2m_{1} (\Delta E_{C} (z)-E)/\hbar ^{2} ]^{{1 \mathord{\left/{\vphantom{1 2}}\right.\kern-\nulldelimiterspace} 2} } 
\label{eq2.16}
\end{eqnarray}
are provided by a piecewise continuous function, which is obtained by dividing each RTS layer by points $z_{p_{l} } =l(z_{p} -z_{p-1} )/2M,\, \, \, p=1...N,\, \, z_{0} =0$, where $M$  is the number of partitions selected $p$-th nanosystem layer, $Ai\, z,\, \, Bi\, z$  are the Airy functions.

The discrete spectrum of an electron $E_{n}$  is determined from the dispersion equation, which in turn is obtained from the boundary conditions for the wave functions and for their probability flows at the RTS heterointerfaces:
\begin{align}
\Psi ^{(p)} (\, E,\, \, z_{p} )=\Psi ^{(p+1)} (E,\, \, z_{p} );\left. \, \, \frac{\rd\Psi _{n}^{(p)} (E,\, \, z)}{m(z)\rd z} \right|_{z=z_{p} -\varepsilon } =\left. \frac{\rd\Psi _{n}^{(p+1)} (E,\, \, z)}{m(z)\rd z} \right|_{z=z_{p} +\varepsilon } \, ,\, \, \varepsilon \to 0.
\end{align}
Besides, using conditions (\ref{eq2.16}) and the normalization condition for the wave function
\begin{align}
\int _{-\infty }^{+\infty }\Psi _{n} (E_{n\bar{k}} ,z)\Psi _{n'}^{*} (E_{n'\bar{k'}} ,z) \rd z=\delta _{nn'}\delta _{kk'}
\end{align}
all unknown coefficients $A^{(0)} ,\, \, B^{(N+1)} ,\, \, A^{(p_{l} )} ,\, \, \, B^{(p_{l} )}$ are found, which completely determine the wave function of the electron $\Psi _{n} (E_{n\bar{k}} ,z)$.

Then, for stationary electronic states within the effective RTS potential $V(z)$, a transition is made from the coordinate representation of the electron Hamiltonian in equation (\ref{eq2.3}) to the representation of the second quantization with a quantized wave function, which is defined as follows:
\begin{align}
\hat{\Psi }(x,y,z)=\sum _{n\, \bar{k}}\Psi _{n\, \bar{k}} (\bar{r}_{0} ,z) \hat{a}_{n\, \bar{k}} =\sum _{\bar{k}}\sum _{n}\Psi _{n\, \bar{k}} (\bar{r}_{0} ,z)  \hat{a}_{n\, \bar{k}}\,,
\end{align}
we obtain the Hamiltonian of noninteracting electrons in the form:
\begin{align}
\hat{H}_{e} =\sum _{n\, ,\bar{k}}E_{n\, \bar{k}} \hat{a}_{n\, \bar{k}}^{+} \hat{a}_{n\, \bar{k}}\,,
\end{align}
where $E_{n\, \bar{k}}$  is determined by relation (\ref{eq2.5}), and the fermionic creation ($\hat{a}_{n\, \bar{k}}^{+}$) and annihilation  ($\hat{a}_{n\, \bar{k}}$) operators of stationary electronic states satisfy the well-known anticommutative relations.

\section{Theory of acoustic phonons modes in plane nitride-based nanostructure}

The spectrum and modes of acoustic phonons in the investigated multilayer nano-RTS is obtained by finding solutions of the equation of motion for the nanostructure elastic medium:
\begin{align}
\rho (z)\frac{\partial ^{2} u_{l} (\bar{r},t)}{\partial t^{2} } =\frac{\partial \sigma _{lk} (\bar{r})}{\partial x_{k} } ;\quad l,\, \, k=(1;\, \, 2;\, \, 3),
\label{eq3.1}
\end{align}
where $x_{1} =x$; $x_{2} =y;$ $x_{3} =z$, $u_{l} =u_{l} (x,y,z,\, t)$  is the component of the elastic displacement vector at the point $(x_{1} ,x_{2} ,x_{3} )=(x,y,z)$ for a point of time $t$, $ \displaystyle \sigma _{ik} (r)=\frac{1}{2} C_{iklm} (z)\left(\frac{\partial u_{l} (r)}{\partial x_{m} } +\frac{\partial u_{m} (r)}{\partial x_{l} } \right),\, \, l,\, \, m=(1;\, \, 2;\, \, 3)$  is the stress tensor,
\begin{align}
\left(\begin{array}{l} {\rho (z)} \\ {C_{iklm} (z)} \end{array}\right)=\sum _{p=0}^{N}\left(\begin{array}{l} {\rho ^{(p)} } \\ {C_{iklm}^{(p)} } \end{array}\right)\left[\theta (z-z_{p-1} )-\theta (z-z_{p+1} )\right] ,\, \, z_{-1} =-\infty ,\, \, z_{N+1} =+\infty
\label{eq3.2} \end{align}
are correspondingly the density $\rho (z)$ and elastic constants $C_{iklm} (z)$ of the nanosystem material, depending on the coordinate $z$.

Since semiconductors $ \rm AlN$ and $\rm GaN$  are of the wurtzite type crystal structure, taking into account the explicit form of the tensor of elastic constants in the Voigt representation ($C_{\alpha\beta}$), we  seek for solutions of equation (\ref{eq3.1}) in the form:
\begin{align}
u_{l} (r,t)=\sum _{p=1}^{N}\left[ {u_{1}^{(p)} (z)} {u_{2}^{(p)} (z)}  {u_{3}^{(p)} (z)} \right]^{T}\left[\theta (z_{p} -z_{p-1} )-\theta (z_{p} -z_{p+1} )\right] \re^{\ri(\omega t-qx)}.
\label{eq3.3}
\end{align}
Using (\ref{eq3.3}) in the equation (\ref{eq3.1}), taking into account (\ref{eq3.2}), it is split into three equations describing all of the acoustic phonons types that occur within an arbitrary $p$-th layer of the RTS:
\begin{align}
-\frac{\rd^{2} u_{1}^{(p)} (z)}{\rd z^{2} } +\ri qc_{1}^{(p)} \frac{\rd u_{3}^{(p)} (z)}{\rd z} -k_{1}^{2} u_{1}^{(p)} (z)=0;\, \, c_{1}^{(p)} &=\frac{C_{13}^{(p)} +C_{44}^{(p)} }{C_{44}^{(p)} } ;\, \, k_{1} =\sqrt{\frac{\rho ^{(p)} \omega ^{2} -q^{2} C_{11}^{(p)} }{C_{44}^{(p)} } }\,,
\label{eq3.4}\\
-\frac{\rd^{2} u_{2}^{(p)} (z)}{\rd z^{2} } +\chi _{2}^2 u_{2}^{(p)} (z)=0;\, \, \, \chi_{2}&=\sqrt{\frac{q^{2} C_{66}^{(p)} -\rho ^{(p)} \omega ^{2} }{C_{44}^{(p)} } }\,,
\label{eq3.5}\\
-\frac{\rd^{2} u_{3}^{(p)} (z)}{\rd z^{2} } +\ri qc_{3}^{(p)} \frac{\rd u_{1}^{(p)} (z)}{\rd z} -k_{3}^{2} u_{3}^{(p)} (z)=0;\, \, \, c_{3}^{(p)} &=\frac{C_{13}^{(p)} +C_{44}^{(p)} }{C_{33}^{(p)} } ;\, \, k_{3} =\sqrt{\frac{\rho ^{(p)} \omega ^{2} -q^{2} C_{44}^{(p)} }{C_{33}^{(p)} } } .
\label{eq3.6}
\end{align}

The solutions of equation (\ref{eq3.5}) describing the shear (SH) acoustic phonons look as follows:
%
\begin{align}
u_{2} (z)&=u_{2}^{(0)} (z)\theta (-z)+\sum _{p=1}^{N}u_{2}^{(p)} (z) \left[\theta (z-z_{p-1} )-\theta (z-z_{p} )\right]+u_{2}^{(N+1)} (z)\theta (z-z_{N}) \nonumber\\
&=B_{2}^{(0)} \re^{\chi _{2}^{(0)} z} \theta (-z)+\sum _{p=1}^{N}\left[A_{2}^{(p)} \re^{-\chi _{2}^{(p)} (z-z_{p-1} )} +B_{2}^{(p)} \re^{\chi _{2}^{(p)} (z-z_{p-1} )} \right] \left[\theta (z-z_{p-1} )-\theta (z-z_{p} )\right] \nonumber\\ 
&+A_{2}^{(N+1)} \re^{-\chi _{2}^{(4)} (z-z_{N} )} \theta (z-z_{N} ).
 \label{eq3.7}
\end{align}

The solutions of equations (\ref{eq3.4}) and (\ref{eq3.6}), which form a system relatively to components $u_1(z)$ and $u_3(z)$, are determined using the following considerations. This system of equations is reduced to a single matrix differential equation having a biquadratic equation as its characteristic equation. Proper functions of such a problem are obtained by applying the Cayley–Hamilton theorem. Finally, the solutions are as follows:
\begin{align}
u_{1(3)} (z)&=u_{_{1(3)} }^{(0)} (z)\theta (-z)+u_{_{1(3)} }^{(N+1)} (z)\theta (z-z_{N} )+\displaystyle \sum _{p=1}^{N}u_{_{1(3)} }^{(p)} (z) \left[\theta (z-z_{p-1} )-\theta (z-z_{p} )\right]; 
\nonumber\\  
u_{1}^{(p)} (z)&=-\ri qc_{1}^{(p)} \left(\frac{\lambda _{1}^{(p)} }{\left\| U_{1}^{(p)} \right\| } \left(A_{1}^{(p)} \re^{\lambda _{1}^{(p)} z} -C_{1}^{(p)} \re^{-\lambda _{1}^{(p)} z} \right)+\frac{\lambda _{2}^{(p)} }{\left\| U_{2}^{(p)} \right\| } \left(B_{1}^{(p)} \re^{\lambda _{2}^{(p)} z} -D_{1}^{(p)} \re^{-\lambda _{2}^{(p)} z} \right)\right);\nonumber\\
 u_{3}^{(p)} (z)&=-\frac{\left(\lambda _{1}^{(p)} \right)^{2} +k_{3}^{2} }{\left\| U_{1}^{(p)} \right\| } \left(A_{1}^{(p)} \re^{\lambda _{1}^{(p)} z} +C_{1}^{(p)} \re^{-\lambda _{1}^{(p)} z} \right)-\frac{\left(\lambda _{2}^{(p)} \right)^{2} +k_{3}^{2} }{\left\| U_{2}^{(p)} \right\| } \left(B_{1}^{(p)} \re^{\lambda _{2}^{(p)} z} +D_{1}^{(p)} \re^{-\lambda _{2}^{(p)} z} \right); 
 \label{eq3.8}
 \end{align}
where $\left\| U_{n}^{(p)} \right\| =\sqrt{q^{2} \left(\lambda _{n}^{(p)} \right)^{2} c_{1}^{2} +\left[\left(\lambda _{n}^{(p)} \right)^{2} +k_{3}^{2} \right]^{2} } ,\, \, n=1,\, 2$ and the roots $\lambda _{n}^{(p)}$ are determined from the relation:
\begin{align}
\lambda _{1,2,3,4}^{(p)} =\lambda _{1,2,3,4}^{(p)} (q,\omega )&=\pm \left[-\frac{q^{2} \left[\left(C_{13}^{(p)} \right)^{2} +2C_{13}^{(p)} C_{44}^{(p)} -C_{11}^{(p)} C_{13}^{(p)} \right]+(C_{33}^{(p)} +C_{44}^{(p)} )^{2} \omega ^{2} }{2C_{33}^{(p)} C_{44}^{(p)} }  \right.\nonumber\\
&\pm \left. \left\{\left[\frac{q^{2} \left[\left(C_{13}^{(p)} \right)^{2} +2C_{13}^{(p)} C_{44}^{(p)} -C_{11}^{(p)} C_{13}^{(p)} \right]+(C_{33}^{(p)} +C_{44}^{(p)} )^{2} \omega ^{2} }{2C_{33}^{(p)} C_{44}^{(p)} } \right]^{2}  \right. \right.\nonumber\\
&-\left.\left.\frac{(\rho ^{(p)} \omega ^{2} -q^{2} C_{11}^{(p)} )(\rho ^{(p)} \omega ^{2} -q^{2} C_{44}^{(p)} )}{C_{33}^{(p)} C_{44}^{(p)} } \right\}^{{1 / 2} } \right]^{{1 / 2} } , 
\nonumber\\  
&\lambda_{1}^{(p)} =-\lambda_{3}^{(p)} ;\, \, \lambda_{2}^{(p)} =-\lambda_{4}^{(p)} .
\label{eq3.9}
\end{align}

In the expressions (\ref{eq3.7}), (\ref{eq3.8}) it is taken into account that $A_{2}^{(0)} =\, B_{2}^{(N+1)} =0$ and $C_{1}^{(0)} =D_{1}^{(0)} =A_{1}^{(N+1)} =B_{1}^{(N+1)} =0$, which is a consequence of ensuring the fulfillment of the conditions that the values of the elastic displacement components cannot grow infinitely in the external semiconductor medium in which the RTS is located, i.e.,
\begin{align}
\left. u{}_{l=1,2,3} (z)\right|_{z\to \pm \infty } \to 0.
\end{align}
Consistently using the boundary conditions for the components $u_{2} (z)$ and components \\ $\displaystyle \sigma _{yz} (z)=\frac{1}{2} C_{44} \frac{\rd u_{2} (z)}{\rd z} \, \re^{\ri(\omega t-qx)}$  of the stress tensor for solutions of (\ref{eq3.7}) in adjacent layers of the RTS:
\begin{align}
\displaystyle \left[\begin{array}{l} {\left. u_{2}^{(p)} (z)\right|_{z=z_{p} -\varepsilon } =\left. u_{2}^{(p+1)} (z)\right|_{z=z_{p} +\varepsilon } ;\, \, } \\ \\ \displaystyle{\left. \sigma _{yz}^{(p)} (z)\right|_{z=z_{p} -\varepsilon } =\left. \sigma _{yz}^{(p+1)} (z)\right|_{z=z_{p} +\varepsilon } } \end{array}\right.
\label{eq3.11}
\end{align}
the dispersion equation is obtained for determining the spectrum $\Omega ^{({\rm SH})} (q)$ of shear acoustic phonons.
Similarly, using the boundary conditions for the components $u_{1(3)} (z)$ and components of the stress tensor $ \sigma _{xz} (z)=\frac{1}{2} C_{44} \left[-\ri qu_{3} (z)+\frac{\rd u_{1} (z)}{\rd z} \right]\, \re^{\ri(\omega t-qx)}$ $\,\,$ and $ \sigma _{zz} (z)=\left[-\ri qC_{13} u_{1} (z)+C_{33} \frac{\rd u_{3} (z)}{\rd z} \right]\re^{\ri(\omega t-qx)}$:
\begin{align}
\left[\begin{array}{l} \displaystyle {\left. u_{1(3)}^{(p)} (z)\right|_{z=z_{p} -\varepsilon } =\left. u_{1(3)}^{(p+1)} (z)\right|_{z=z_{p} +\varepsilon } } \\ \\ \displaystyle {\left. \sigma _{xz}^{(p)} (z)\right|_{z=z_{p} -\varepsilon } =\left. \sigma _{xz}^{(p+1)} (z)\right|_{z=z_{p} +\varepsilon } ;\, \, \left. \sigma _{zz}^{(p)} (z)\right|_{z=z_{p} -\varepsilon } =\left. \sigma _{zz}^{(p+1)} (z)\right|_{z=z_{p} +\varepsilon } } \end{array}\right.
\label{eq3.12}
\end{align}
the dispersion equation is obtained, from which the mixed spectrum $\Omega {}^{({\rm FL,DL})} (q)$ of flexural (FL) and dilatational (DL) phonon modes is found. These acoustic phonon modes are determined using components $u_{1(3)} (z)$ as: $u^{(FL)} (z)=u^{(FL)} \left[u_{1}^{(A)} (z),u_{3}^{(S)} (z)\right]$ and $u^{(DL)} (z)=u^{(SL)} \left[u_{1}^{(S)} (z),u_{3}^{(A)} (z)\right]$, where the indices ``$S$'' and ``$A$'' are used to mark the symmetric and antisymmetric functions of $z$, correspondingly \cite{Pokatilov2003,Boyko2020,Boyko2020-1}.
\section{Theory of electron-acoustic phonon interaction in the plane AlN/GaN nanostructure}
Using the boundary conditions (\ref{eq3.11}) and (\ref{eq3.12}), the coefficients $A_{2}^{(p)} ,\, \, B_{2}^{(p)}$  and $A_{1}^{(p)} ,\, \, B_{1}^{(p)} ,\, \, C_{1}^{(p)} ,\, \, D_{1}^{(p)}$  in solutions (\ref{eq3.7}) and (\ref{eq3.8}), correspondingly, can be expressed through one of them being found from the normalization condition for shear phonons
\begin{align}
\int _{-\infty }^{+\infty }\rho (z)u_{2} (q,\omega ,z)u_{2}^{*} (q',\omega ,z)\rd z= \frac{\hbar }{2l_{x} l_{y} \omega } \delta _{qq'}
\end{align}
and from normalization condition for dilatational and flexural phonons \cite{Boyko2020-1,Stroscio2001}:
\begin{align}
\int _{-\infty }^{+\infty }\rho (z)\left[u_{1} (q,\omega ,z)u_{1}^{*} (q',\omega ,z)+u_{3} (q,\omega ,z)u_{3}^{*} (q',\omega ,z)\right]\rd z= \frac{\hbar }{2l_{x} l_{y} \omega } \delta _{qq'}
\end{align}
where the values $l_{x}$  and $l_{y}$  provide the geometric dimensions of the RTS cross-section area by plane $xOy$.

While quantizing the field of elastic displacements using the well-known quantum mechanical method~\cite{Stroscio2001,Tkach2003}, the components for the Hamiltonian of acoustic phonons are obtained in the canonical form of the second quantization representation, that is, in the form of the sum of two components for shear and flexural and dilatational phonons:
\begin{align}
\hat{H}_{{\rm ac}} &=\hat{H}_{{\rm ac}}^{{\rm (FL,DL)}} +\hat{H}_{{\rm ac}}^{{\rm (SH)}} =\sum _{n_{1} }\Omega _{n_{1} }^{{\rm (FL,DL) }} (q)\left[\hat{b}_{n_{1} }^{+} (q)\hat{b}_{n_{1} } (q)+\frac{1}{2} \right] \nonumber\\
&+\sum _{n_{2} }\Omega _{n_{2} }^{({\rm SH})} (q)\left[\hat{b}_{n_{2} }^{+} (q)\hat{b}_{n_{2} } (q)+\frac{1}{2} \right],
\end{align}
where $\hat{b}_{n}^{+} (q)$ and $\hat{b}_{n} (q)$  are the boson phonon state creation and annihilation operators, correspondingly.

Taking into account the normalization conditions, as well as relation (\ref{eq3.2}), the expression for the elastic displacement operator is obtained using Fourier transform of $u(q,\omega,z)$, which can be represented as follows:
\begin{align}
\hat{u}(q,\omega ,R)&=\sum _{p=0}^{N}\sum _{q,\tilde{n}}\sqrt{\frac{\hbar }{2l_{x} l_{y} \rho ^{(p)} \omega ^{(\beta )} } }  \left[ \hat{b}_{\tilde{n}} (q)+\hat{b}_{\tilde{n}}^{+} (-q)\right] \, w_{\tilde{n},l}^{(p)} (q,\omega ,z)\re^{\ri\bar{q}\bar{R}} \nonumber\\
&\times\left[\theta (z-z_{p-1} )-\theta (z-z_{p+1} )\right]; \nonumber\\
&w_{\tilde{n},l}^{(p)} (q,\omega ,z)=\sqrt{\rho ^{(p)} } u_{\tilde{n},l}^{(p)} (q,\omega ,z); \quad\tilde{n}=\{n_{1},n_{2}\},\, \, \bar{q}\|\bar{R}, \, \, \beta =\{ ({\rm SH}),({\rm FL,}\, {\rm DL})\} .
\label{eq4.4}
\end{align}

In semiconductors with the wurtzite crystal structure, the displacement of the conduction band is not determined by a single constant of the deformation potential as in \cite{Stroscio2001}, and looks like \cite{Yan2014}:
\begin{align}
\Delta E_{C } =a_{1} \varepsilon _{zz} +a_{2} \varepsilon _{\bot }\,,
\end{align}
where $\varepsilon _{\bot } =\varepsilon _{xx} +\varepsilon _{yy}$ and $\varepsilon _{zz}$  are strain tensor components, $a_{1} =a_{1c} -D_{1} ,\, \, a_{2} =a_{2c} -D_{2}$, $a_{1c} ,\, \, a_{2c} ,\, \, D_{1} ,\, \, D_{2}$  are the deformation potential constants  \cite{Yan2014,Vurgaftman2001}.

Then, the interaction Hamiltonian due to the deformation potential in the representation  of the second quantization  in terms of phonon variables is defined as follows:
\begin{align}
\hat{H}_{{\rm def}} =\hat{H}_{{\rm def}}^{{\rm (FL,DL)}},
\end{align}
where the Hamiltonian components for  flexural and dilatational phonons are:
\begin{align}
 \hat{H}_{{\rm def}}^{{\rm (FL,DL)}} &=\sqrt{\frac{\hbar }{2l_{x} l_{y} } } \sum _{q\, n_{1} }\sum _{p=0}^{N}\frac{1}{\sqrt{\rho ^{(p)} \omega _{n_{{\rm 1}} }^{{\rm (FL,DL)}} } } [\hat{b}_{n_{1} } (q)+\hat{b}_{n_{1} }^{+} (-q)]  \left[\ri q(a_{2c} -D_{2} )u_{1}^{(p)} (q,\omega _{n_{{\rm 1}} }^{{\rm (FL,DL)}} ,z)+\right. \nonumber\\
&\left. +(a_{1c} -D_{1} )\frac{\partial u_{3}^{(p)} (q,\omega _{n_{{\rm 1}} }^{{\rm (FL,DL)}} ,z)}{\partial z} \right]\left[\theta (z -z_{p-1} )-\theta (z -z_{p+1} )\right]\re^{\ri\bar{q}\cdot \bar{R}}, 
\label{eq4.7}
\end{align}
where frequency $\omega _{n_{{\rm 1}} }^{{\rm (FL,DL)}}$ refers to the spectrum of these phonons, $n_{1}$ is the number of their spectrum level.

In the Hamiltonian (\ref{eq4.7}), there is no contribution from the shear acoustic phonons, since  
\[\partial [\bar{w}_{n_{2},2}^{(p)} (q,\omega ,z)\re^{\ri\bar{q}\cdot \bar{R}} ]/\partial y=[\bar{q}\cdot \bar{w}_{n_{2},2}^{(p)} (q,\omega ,z)]\re^{\ri\bar{q}\cdot \bar{R}} =0,\, \, \bar{q}\bot \bar{w}_{n_{2},2} (q,\omega ,z)),\] 
similarly to that determined in the framework of the single-well nanostructure model in the paper \cite{Pokatilov2004} and it has not been properly treated in the paper \cite{Yang2013}.

Now, the Hamiltonian describing the interaction of electrons with acoustic phonons looks as follows:
\begin{align}
\hat{H}_{e-{\rm def}} =\sum _{n,\, \, n' ,\, n_{1} ,\, \bar{k},\, \bar{q}}\sum _{p=0}^{N} F_{nn_{1} } (q)\hat{a}_{n' ,\, \, \bar{k}+\bar{q}}^{+} \hat{a}_{n\, \bar{k}}  \left[\hat{b}_{n_{1} } (q)+\hat{b}_{n_{1} }^{+} (-q)\right]\left[\theta (z -z_{p-1} )-\theta (z -z_{p+1} )\right],
\label{eq4.8}
\end{align}
where
\begin{align}
F_{nn_{1} n'} (q)&=\sqrt{\frac{\hbar }{2l_{x} l_{y} \rho ^{(p)} \omega _{n_{{\rm 1}} }^{{\rm (FL,DL)}} } } \int _{z_{p-1} }^{z_{p} }\Psi ^{(p)} (E_{n} ,z)\left[\ri q(a_{2A} -D_{2} )u_{1}^{(p)} (q,\omega _{n_{{\rm 1}} }^{{\rm (FL,DL)}} ,z)\right.   \nonumber\\  
&\left. +(a_{1c} -D_{1} )\frac{\partial u_{3}^{(p)} (q,\omega _{n_{{\rm 1}} }^{{\rm (FL,DL)}} ,z)}{\partial z} \right]\re^{\ri\bar{q}\bar{r}} \Psi _{n'}^{*(p)} (E_{n} ,z)\rd z.
\label{eq4.9}
\end{align}

Finally, the Hamiltonian for a system of an electron with acoustic phonons in a multilayer RTS:
\begin{align}
\hat{H}=\hat{H}_{e} +\hat{H}_{ac} +\hat{H}_{e-{\rm def}}.
\end{align}

Taking into account the fact that the electronic spectrum of the investigated RTS contains only a discrete component, for its renormalization by interaction with acoustic phonons, it is necessary to carry out the Fourier transform of the Green's function, which satisfies the Dyson equation \cite{Tkach2003}:
\begin{align}
G_{n} (\Omega )=\left[\Omega -E_{n\bar{k}} -M_{n} (\Omega )\right]^{-1}.
\label{eq4.11}
\end{align}

The mass operator in the Dyson equation for one-phonon approximation ($\eta\rightarrow\pm0$) is defined as follows:
\begin{align}
\begin{array}{l} \displaystyle M_{n} (\Omega ,\bar{k})=\sum _{qn_{1} n'}\left|F_{nn_{1} n'} (q)\right|^{2} \left[\frac{1+\nu _{n_{1}\bar{q} }^{({\rm FL,DL})} }{\Omega -E_{n',\bar{k}+\bar{q}} -\Omega _{n_{{\rm 1}}\bar{q} }^{{\rm (FL,DL)}} +\ri\eta } +\frac{\nu _{n_{1} \bar{q}}^{({\rm FL,DL})} }{\Omega -E_{n',\bar{k}+\bar{q}} +\Omega _{n_{{\rm 1}}\bar{q} }^{{\rm (FL,DL)}} +\ri\eta } \right], \end{array}
\label{eq4.12}
\end{align}
where $\Omega _{n_{{\rm 1}}\bar{q} }^{{\rm (FL,DL)}} =\hbar \omega _{n_{{\rm 1}} }^{{\rm (FL,DL)}}(q)$ and $\nu _{n_{1} \bar{q}}^{({\rm FL,DL})} =(\re^{\hbar \omega _{n_{{\rm 1}} }^{{\rm (FL,DL)}}(q) /kT} -1)^{-1}$  are the average occupation numbers determined by the flexural and dilatational acoustic phonon modes.

Using the mass operator (\ref{eq4.12}),  the mechanisms of the interaction of electrons with acoustic phonons influence are studied, i.e., the shift of the electronic spectrum ($\Delta _{n}$) and the decay rate of the electronic state ($\Gamma _{n}$).

Having tacking into account that in QCD the movement of electrons occurs in the direction of the axis $Oz$, that is, perpendicular to the RTS layers, then in direct calculations $\bar{k}=0$ it should be taken, then according to (\ref{eq2.5}): $\Omega =E_{n}$. Then, the renormalized energy $\tilde{E}_{n}$ of the electronic level is determined by the pole of the Fourier transform of the Green's function (\ref{eq4.11}) taking into account (\ref{eq4.9}), which is similar to finding solutions of the dispersion equation:
\begin{align}
E_{n} -E_{n,\bar{q}} -M_{n} (\Omega )=0
\end{align}
here,
\begin{align}
\Delta _{n}& ={\Re} M_{n} (\Omega =E_{n} ,\bar{k}=0)\nonumber\\ 
&=\frac{l_{x} l_{y} }{(2\piup )^{2} } \sum _{n_{1} }\left(\nu _{n_{1} }^{({\rm FL,DL})} +\frac{1}{2} \pm \frac{1}{2} \right)\iint \nolimits _{}{\mathscr{P}} \left(E_{n} -E_{n,\bar{q}} -\Omega _{n_{{\rm 1}} }^{{\rm (FL,DL)}} \right)^{-1} \left|F_{nn_{1} n'} (q)\right|^{2} d^{2} q  ;\,  \nonumber\\ 
\Gamma _{n} &=-2 { \Im} M_{n} (\Omega =E_{n} ,\bar{k}=0)\nonumber\\ 
&=\frac{l_{x} l_{y} }{2\piup } \sum _{n_{1} }\left(\nu _{n_{1} }^{({\rm FL,DL})} +\frac{1}{2} \pm \frac{1}{2} \right)\iint \nolimits _{}\delta \left(E_{n} -E_{n,\bar{q}} -\Omega _{n_{{\rm 1}} }^{{\rm (FL,DL)}} \right)^{-1} \left|F_{nn_{1} n'} (q)\right|^{2} d^{2} q \,,
\label{eq4.14}
\end{align}
where in relations (\ref{eq4.14}), the symbol $\mathscr{P}$  means that the integral is taken via the Cauchy principal value.

\begin{table}[htb]
\caption{Physical parameters of GaN and AlN wurtzite semiconductors.}
\label{tbl-smp1}
\begin{center}
\renewcommand{\arraystretch}{0}
\begin{tabular}{|c||c|c|c|c|c|c|c|}
\hline
&$\rho\,\,(\text{kg/m}^{3})$&$C_{11}\left(\rm GPa \right)$&$C_{12}\left(\rm GPa \right)$&$C_{13}\left(\rm GPa \right)$&$C_{33}\left(\rm GPa \right)$&$C_{44}\left(\rm GPa \right)$&$C_{66}\left(\rm GPa \right)$\strut\\
\hline
\rule{0pt}{2pt}&&&&&&&\\
\hline
\raisebox{-1.7ex}[0pt][0pt]{}
       $\rm GaN$& 6150&  390&  145&  106&  398&  105&  123\strut\\
\cline{1-8}
      $\rm AlN$& 3255&  396&  137&  108&  373&  116&  130  \strut\\
\hline
&$ m/m_{e}$&$\varepsilon$&$P_{Sp}\left(\rm C/m^{2} \right)$&$a_{1c}\left(\rm eV \right)$&$a_{2c}\left(\rm eV \right)$&$D_{1}\left(\rm eV \right)$&$D_{2}\left(\rm eV \right)$\strut\\
\hline
\rule{0pt}{2pt}&&&&&&&\\
\hline
\raisebox{-1.7ex}[0pt][0pt]{}
       $\rm GaN$& 0.186&  10&  $-0.034$&  $-6.5$&  $-11.8$&  $-3.0$&  3.6\strut\\
\cline{1-8}
      $\rm AlN$& 0.322&  8.5& $ -0.081$&  $-9.0$&  $-9.0$&  $-3.1$&  3.8 \strut\\
\hline
\end{tabular}
\renewcommand{\arraystretch}{1}
\end{center}
\end{table}

\begin{figure}[!b]
\centerline{\includegraphics[width=0.53\textwidth]{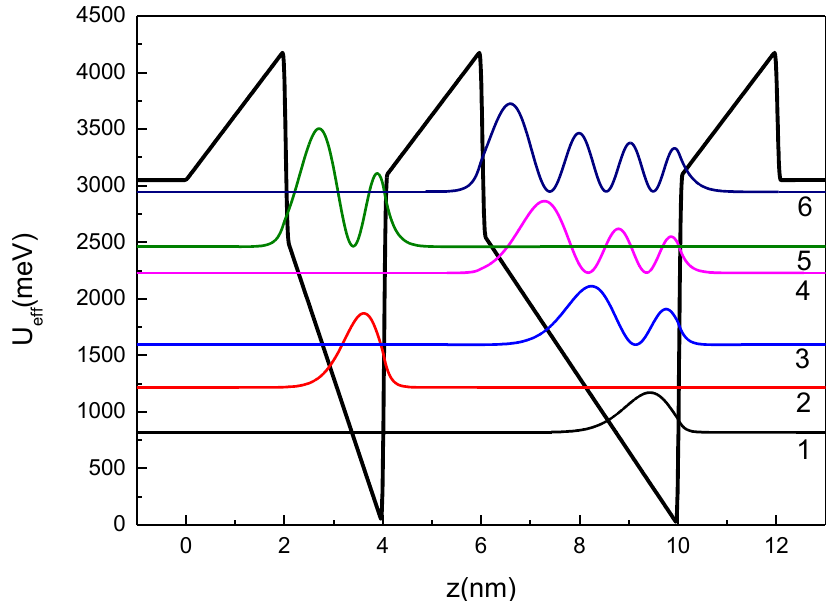}}
\caption{(Colour online) Dependences of the effective potential $U_{\text{eff}}(z)$ and squared moduli $\left|\Psi (E_{n} ,z)\right|^{2}$ of the wave functions for the first $n=1,2,\ldots,6$ stationary states of an electron.} \label{fig-smp2}
\end{figure}

Then, the complete shift of the stationary electronic spectrum $n$-th energy level due to the flexural and dilatational acoustic phonons $\Delta _{n} =\Delta _{n}^{{\rm (FL,}\, {\rm DL)}}$, which gives the renormalized energy value:
\begin{align}
\tilde{E}_{n} =E_{n} +\Delta _{n}.
\end{align}

\begin{figure}[!b]
\centerline{\includegraphics[width=0.90\textwidth]{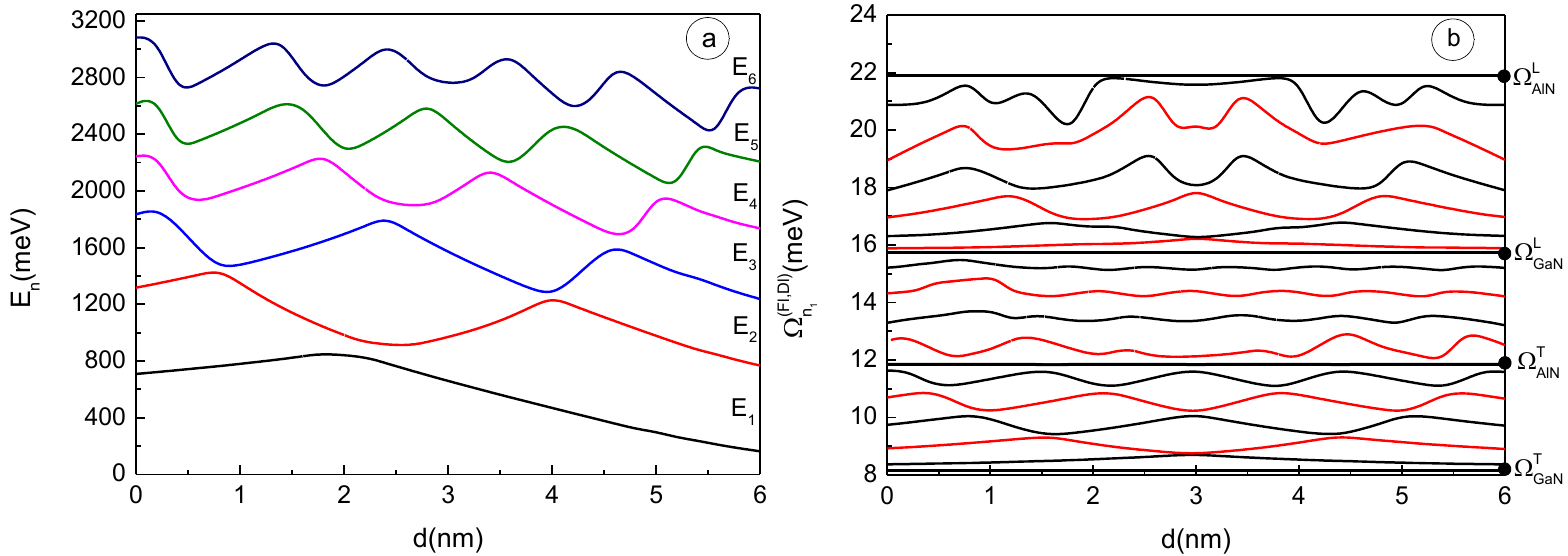}}
\caption{(Colour online) Dependences of the stationary electronic states spectrum ($E_{n}, n=1,2,\ldots,6$) (a), as well as the spectrum of dilatational-flexural acoustic phonons ($\Omega _{n_{1} }^{(\rm FL,DL)}$) (b), calculated at $q=24/(\Delta_{1}+\Delta_{2}+\Delta_{3}+d_{1}+d_{2})$, on the position ($0\leqslant d\leqslant d_{1}+d_{2}$) of the internal potential barrier in the total potential well.} \label{fig-smp3}
\end{figure}

\begin{figure}[!b]
\centerline{\includegraphics[width=0.92\textwidth]{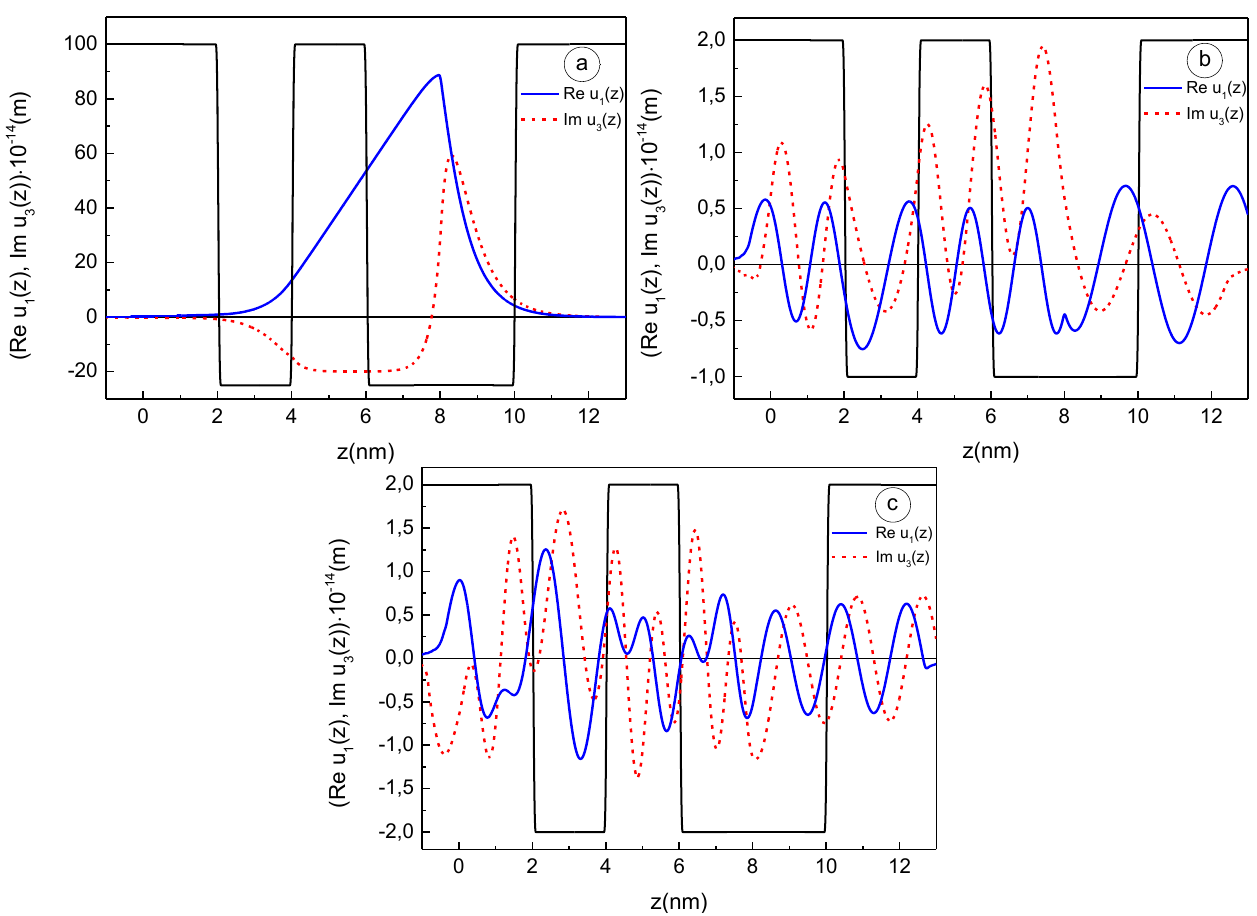}}
\caption{Dependences of the displacement field components on $z$, calculated at $q=24/(\Delta_{1}+\Delta_{2}+\Delta_{3}+d_{1}+d_{2})$ for acoustic phonon energy values: $\Omega^{(\rm FL,DL)}=(5.561,\,13.327,\,19.556) \, \rm meV$.} \label{fig-smp4}
\end{figure}
\section{Discussion of the results}

Using the above developed  theory of the interaction of electrons with acoustic phonons in plane nitride nanosystems, the spectrum of these quasiparticles has been calculated, as well as the displacements of the stationary electron spectrum due to this interaction. The mentioned values were calculated using the geometrical parameters of a plane double-well GaN/AlN nanosystem with such geometrical parameters: the thickness of layers corresponding to the potential barriers --- $\Delta _{1} =\Delta _{2} =\Delta _{3} =2\, \, {\rm nm}$, the width of potential wells --- $d_{1} =2\, \, {\rm nm};\, \, \, d_{2} =4 \, {\rm nm}$. The physical parameters of semiconductor materials corresponding to the nanosystem layers environment were taken from the papers \cite{Piprek2007,Yan2014,Vurgaftman2001}. They are presented in the table~1, $m_{e}$  is free electron mass.

In figure~2, the potential profile of the studied nanostructure, calculated at $T=300$~K  is shown. The figure also shows the square moduli of wave functions for electronic states created in the nanosystem by dimensional quantization effect. To present it more clearly, the values of $\left|\Psi (E_{n} ,z)\right|^{2}$ are aligned with corresponding $E_{n}$ values in the energy scale. The effect of the electric field formed by the total polarization value (\ref{eq2.11}), as well as the localization of the electron within the nanosystem for each of the stationary states, is clearly visible from the above figure.
\begin{figure}[!b]
\centerline{\includegraphics[width=1.00\textwidth]{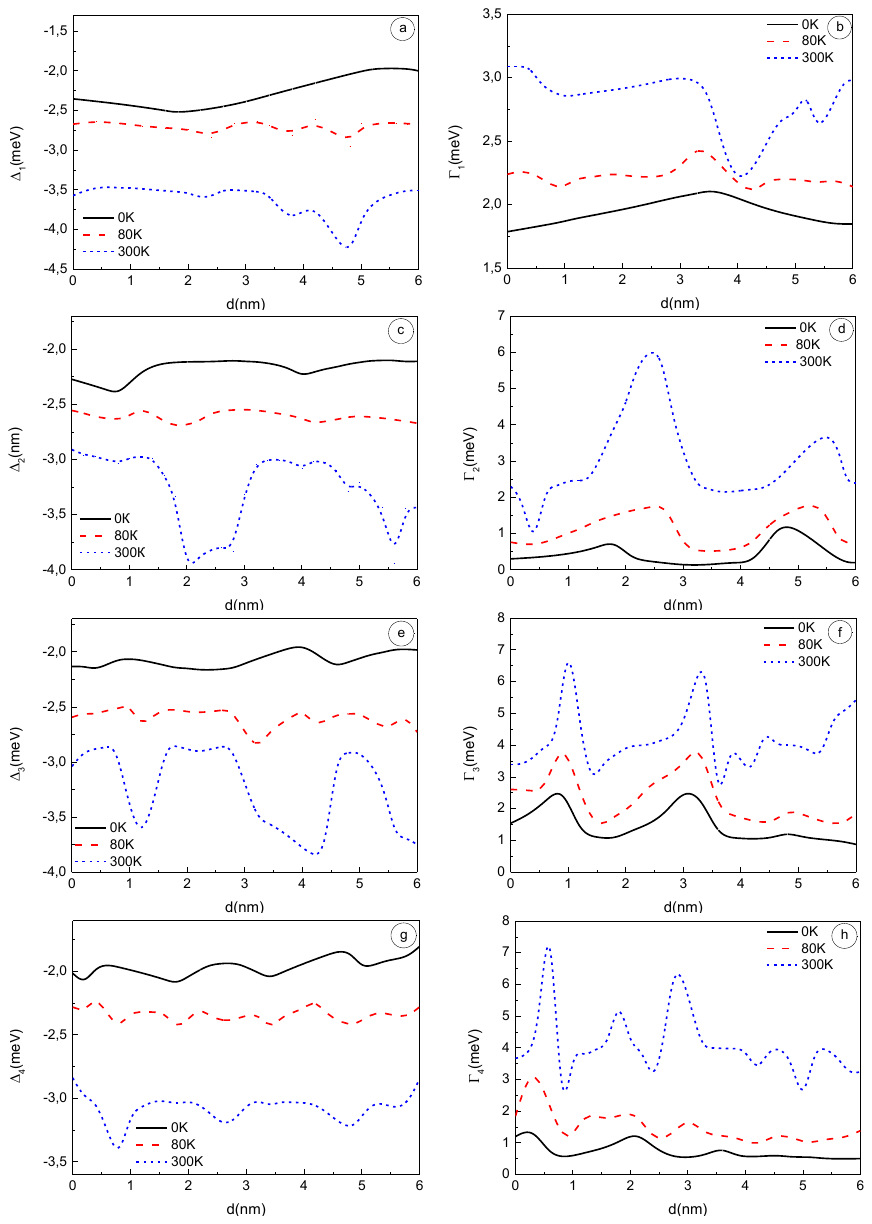}}
\end{figure}
\begin{figure}[h]
\centerline{\includegraphics[width=0.9\textwidth]{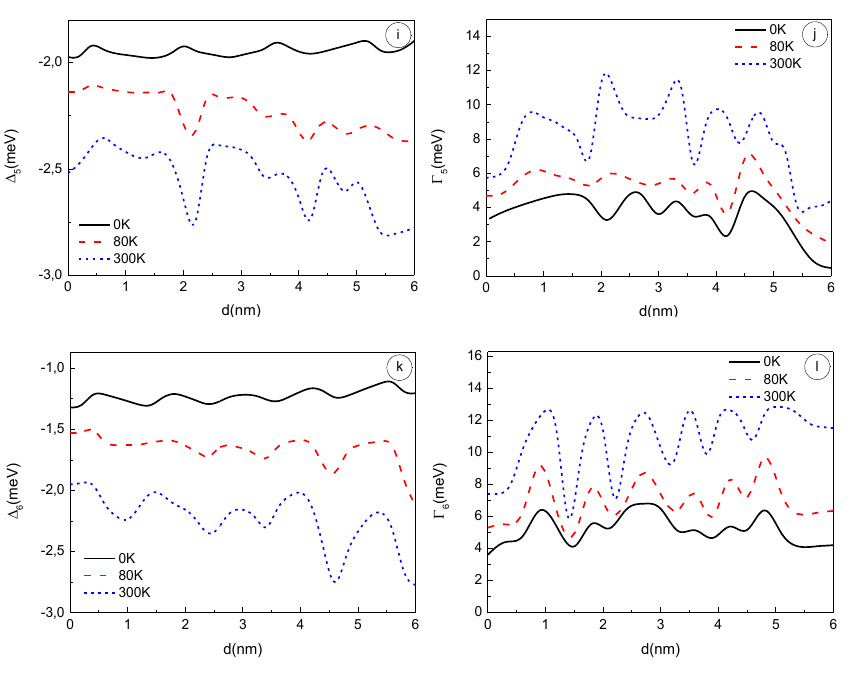}}
\caption{Dependences of the electronic states shifts $\Delta_{n}(d)$ and decay rates $\Gamma_{n}(d)$ calculated at $q=24/(\Delta_{1}+\Delta_{2}+\Delta_{3}+d_{1}+d_{2})$ for temperature $T$: 0~K (solid black line), 80~K (dashed red line), 300~K (short dashed blue line).} \label{fig-smp5}
\end{figure}
In figure~3 (a), (b) dependences calculated at $T = 300$~K,  on the position of the internal potential barrier relatively to the input and to the output potential barriers in the total potential well are presented, i.e., on the value of $d (0\leqslant d\leqslant d_{1}+d_{2})$,  for the stationary electronic spectrum $E_{n}(d)$ [figure 3 (a)] and flexural and dilatational acoustic phonons  spectrum $\Omega _{n_{1} }^{(\rm FL,DL)}(d)$ [figure~3 (b)].

As it can be seen from figure 3 (a), in the dependences $E_{n}(d)$ with an increase of $d$, for each stationary electronic state with number $n$, respectively, $n$ maxima and $n-1$ minimum are formed. In addition, for the energies of the electronic spectrum, the fulfillment of the next condition is provided by  direct calculations:
\begin{align}
\displaystyle \left. E_{n} (d)\right|_{d\to 0} -\left. E_{n} (d)\right|_{d\to d_{1} +d_{2} } \displaystyle \approx \left. \left[ V_{E} (z)+V_{HL} (z)+V_{H} (z)\right] \right|_{z\to d_{1} +d_{2} }^{z\to 0}.
\label{eq5.1}
\end{align}
The dependence (\ref{eq5.1}) is caused by the action of a strong internal electric field, which strongly deforms the potential profile of the nanosystem, and thus is a decisive factor of the action on the electronic spectrum.

As it can be seen from figure 3 (b), the spectrum of dilatational and flexural acoustic phonons $\Omega _{n_{1} }^{(\rm FL,DL)}$ at a fixed value of $q$ is formed within three separate regions limited by energies $\Omega _{\rm GaN}^{\rm T},\, \Omega _{\rm AlN}^{\rm T},\,\Omega _{\rm GaN}^{\rm L},\, \Omega _{\rm AlN}^{\rm L}$, determined respectively by the group propagation velocities of transverse (``T'') and longitudinal (``L'') acoustic waves in bulk materials $\rm GaN$ and $\rm AlN$ \cite{Pokatilov2004,Boyko2020-1}. Each of these regions dependences is characterized by a set of features that are manifested only for this energy range. Thus, for the first region defined as $\Omega _{\rm GaN}^{\rm T}\leqslant\Omega\leqslant \Omega _{\rm AlN}^{\rm T} $, the branches of the spectrum of acoustic phonons  are symmetric relatively to the position of the potential barrier in the total potential well, forming with an increase of $d$, correspondingly, $n_{1}$ maxima and $n_{1}-1$ minima. The two lower branches of the second region, defined as $\Omega _{\rm AlN}^{\rm T}\leqslant\Omega\leqslant \Omega _{\rm GaN}^{\rm L} $, also behave symmetrically relatively to the point $d/2$, though in both of them there are formed six maxima and five minima. In the other two branches of this region, symmetry relatively to the point $d/2$ is broken and seven maxima and six minima are formed, correspondingly. The branches of the third region, defined as $\Omega _{\rm GaN}^{\rm L}\leqslant\Omega\leqslant \Omega _{\rm AlN}^{\rm L} $, behave similarly to the branches of the first region. This is especially true for the first four branches that behave in almost the same way. The last two branches, despite the symmetry relatively to  the point $d/2$, already form six maxima and five minima each.

In figure 4 (a), (b), (c), the components of the elastic displacement field $u_{1}(z)$ and $u_{3}(z)$ calculated at a fixed value of  $q=24/(\Delta_{1}+\Delta_{2}+\Delta_{3}+d_{1}+d_{2})$ are shown. The values of the acoustic phonons energies, used in the calculations, were chosen so that they correspond to each of the spectrum regions established above. Thus, the dependences  shown in figure 3 (a)  correspond to the first region of dependencies in figure 3 (b), dependencies in figure 4 (a) correspond to the second region,  the dependences in figure~4~(c) correspond to the third region.

It can be seen from figure 4 (a), (b), (c) that with an increase of the spectrum number of the branch  $n_{1}$, and accordingly, the energy of acoustic phonons corresponding to these branches, dependences $u_{1}(z)$ and $u_{3}(z)$ tend to increase the number of maxima and minima that are formed by calculated dependences over the given range of $z$. In addition, it should be noted that the effect, which consists of the simultaneous formation of maxima of function $u_{1}(z)$, accordingly, minima of function  $u_{3}(z)$, as it was established in~\cite{Boyko2020,Boyko2020-1}, is mainly observed for the dependences shown in figure 4 (a), (b), (c). However, in this case, the formation of these extrema occurs at arbitrary points inside separate the nanosystem layers and they are a little offset from each other, and are not formed in the middle of these layers, as established in \cite{Boyko2020-1}, where acoustic phonons were studied in nanostructures with identical geometric parameters of potential wells. In the media to the left ($z<0$) and to the right ($z>z_{5}$) of the studied nanosystem, the displacement components $u_{1}(z)$ and $u_{3}(z)$  monotonously decrease according to the relations.

The shifts of the stationary electronic states and decay rates of these energy levels due to the interaction of electrons with acoustic phonons,  calculated on the dependence of the internal potential barrier position in the total potential well of the studied nanostructure, are presented in figure~5. The direct calculations were performed for three different values of temperature $T$: 0~K, 80~K, 300~K. Such a choice of temperature values is taken for the following reasons: the value of 0~K corresponds to the most conventional case, that is, in the above-mentioned papers \cite{Pokatilov2004,Yang2013} calculations were carried out precisely in this approximation; the value of 80~K corresponds to the nanodevice operation using the liquid nitrogen cooling \cite{Lim2017}, the value of 300~K corresponds to nanodevices, that can operate at room temperature \cite{Fujikawa2019,Li2019}.

Before analyzing the dependencies shown in figure 5, it should also be noted that the calculated acoustic phonon energies for wurtzite semiconductors  AlN and GaN should correspond to the first Brillouin zone, that is, they are limited by maximum values of the order of 25--30 meV \cite{Boyko2020}. In this case, the conditions are fulfilled at 0~K:
\begin{align}
E_{nn'} >\Omega _{n_{1} }^{(\rm FL,DL)} ,\, \, \delta \left(E_{n} -E_{n'} -\Omega _{n_{1} \bar{q}}^{(\rm FL,DL)} -\hbar ^{2} \bar{q}^{2} /2m_{n}^{(\rm eff)} \right)\ne 0
\end{align}
whence, due to the properties of the Dirac delta function and the dependencies in figure~3 (a), it follows that the decay rates are nonzero in the entire change range of $d$ for all numbers $n$ of electronic levels ($\Gamma_{n}\ne 0$).

In figure 5 (a), (c), (e), (g), (i), (k), the dependences of the shifts for each of the six energy levels due to the interaction with acoustic phonons at different temperatures are presented. As it can be seen, in the lower boundary of the cryogenic temperature ($T=0$~K), the presented dependences of the energy levels shift  behave similarly to the dependences of the energy spectrum shown in figure 3 (a). However, for the upper boundary of the cryogenic temperature ($T=80$~K), the effects forming additional minima in the dependences $\Delta_{n}(d)$ have already started to appear. Such effects are caused by the complex behavior of the binding functions at a non-zero temperature. In most cases, in the vicinity of the values of $d$, which corresponds to the anti-crossings created by the dependences $E_{n}(d)$ and $E_{n+1}(d)$ of the neighboring energy levels, the displacements of these levels also tend to converge their values. At $T\neq0$, such an effect is no longer general, but it is only partial. The dependences $\Delta_{n}(d)$ calculated at room temperature ($T = 300$~K) show a significant increase of the displacements absolute values, being formed at $T = 80$~K and also a significant deformation of these dependences compared to cryogenic temperatures. It should be noted that all electronic states shifts  are negative, which leads to the displacement of each  energy level  to a lower energy region. It is also seen from figure 5 (a), (c), (e), (g), (i), (k) that with an increase of the energy level number $n$, the absolute values of their shifts decrease, and this property is valid for all temperature values that were used in the calculations. Thus, the interaction of electrons with acoustic phonons leads to a decrease of the generated or detected frequency of the electromagnetic field in the case of QCL and QCD.

Then, in figure 5 (b), (d), (f), (h), (j), (l)  the decay rates $\Gamma_{n}$ dependencies of electronic states on the~$d$ values, calculated at the same temperature values as the electron energy shifts $\Delta_{n}$, are shown. As it can be seen from the above dependencies $\Gamma_{n}(d)$, the decay rates increase rapidly with temperature ($T$) increasing. Moreover, as it can be seen from the dependencies $\Gamma_{n}$ calculated at $T = 0$~K, they form such  number of maxima which is equal to the corresponding number of the electronic level $n$. With an increase of temperature, the following transformation of the $\Gamma_{n}(d)$ dependences takes place: the extrema formed at $T = 0$~K remain, their absolute values being increased in 2--4 times at $T = 300$~K, and their position $d$ changes slightly. In addition, at $T \neq 0$~K, additional extrema appear in the dependencies $\Gamma_{n}(d)$, which is associated with the behavior of the $\rm Im M_{n}$ function. It should also be noted that with a change in $n$ there is no monotonous change in the absolute values of $\Gamma_{n}(d)$, since this was observed in the case of  $\Delta_{n}$ (which decrease with the increase of $n$), and in this case: $\max \left|\Gamma _{2} (d)\right|\approx \max \left|\Gamma _{3} (d)\right|\approx \max \left|\Gamma _{4} (d)\right|,\, \, \max \left|\Gamma _{5} (d)\right|\approx \max \left|\Gamma _{6} (d)\right|$.

\section{Conclusions}

1. An analytical theory of the interaction of electron-acoustic phonons   for  a multilayer nitride-based AlN/GaN resonant tunnelling structure  was developed using the exact solutions for the components of the elastic displacement field for a semiconductor medium and solutions the Schr\"odinger and Poisson system of equations.

2. The dependences of the electronic spectrum and spectrum of acoustic phonons spectrum on the geometric parameters of the nanosystem were investigated.

3. Using the temperature Green's functions method, calculations were performed and the dependences of the electronic states shifts and their decay rates due to the interaction of electrons with acoustic phonons at different temperatures were studied.

4. It has been determined that the electron-phonon interaction leads to a shift of the quantum electronic transitions energies in the nanostructure to a region of lower energies and causes an increase of the decay rates of electronic states.

%
%
\newpage
\ukrainianpart

\title{Взаємодія електронів з акустичними фононами в AlN/GaN резонансно-тунельних наноструктурах за різних температур}

\author{І.В. Бойко, М.Р. Петрик}
\address{Тернопільський національний технічний університет імені Івана Пулюя,\\  вул. Руська, 56, 46001  Тернопіль, Україна}
\makeukrtitle
\begin{abstract}
\tolerance=3000%
 Використовуючи метод температурних функцій Гріна та рівняння Дайсона, вперше розвинена теорія взаємодії електронів з акустичними фононами у багатошарових нітридних AlN/GaN наноструктурах у випадку довільних температур. Отримано складові гамільтоніана, що описують систему електрона з акустичними фононами та величини зміщень спектру електрона, зумовлені електрон-фононною взаємо\-дією. Встановлено залежності рівнів спектру електронів акустичних фононів у залежності від положення внутрішнього потенціального бар’єра у досліджуваній наноструктурі. Виконано розрахунки зміщень та згасань рівнів електронного спектру для різних значень температури $T$.
\keywords акустичний фонон, електрон-фононна взаємодія, функція Гріна, рівняння Дайсона, нітридна наноструктура
\end{abstract}


\begin{thebibliography}{10}
\bibitem{Fujikawa2019} Fujikawa S., Ishiguro T., Wang K., Terashima W., Fujishiro H., Hirayama H., J. Cryst. Growth, 2019, \textbf{510}, 47--49, \doi{10.1016/j.jcrysgro.2018.12.027}.

\bibitem{Li2019} Li J., Wan T., Chen C., Semicond. Sci. Technol., 2019, \textbf{34}, 075018, \doi{10.1088/1361-6641/ab1401}.

\bibitem{Lim2017} Lim C.B., Ajay A., Lahnemann J.,  Bougerol C., Monroy E., Semicond. Sci. Technol., 2017, \textbf{32}, No.~12,
        125002, \doi{10.1088/1361-6641/aa919c}.

\bibitem{Mensz2019} Mensz P.M., Dror B., Ajay A., Bougerol C., Monroy E., Orenstein M., Bahir G., J. Appl. Phys., 2019, \\ \textbf{125}, No.~17, 174505, \doi{10.1063/1.5079408}.

\bibitem{Bernardini1998} Bernardini F., Fiorentini V., Phys. Rev. B, 1998, \textbf{57},
        No.~16,  R9427--R9430, \\ \doi{10.1103/PhysRevB.57.R9427}.
        
\bibitem{Bernardini1999} Bernardini F., Fiorentini V., Phys. Status Solidi B, 1999, \textbf{216},
        No.~1,  391--398,\\ \doi{10.1002/(SICI)1521-3951(199911)216:1<391::AID-PSSB391>3.0.CO;2-K}.
        
\bibitem{Saha2016}  Saha S., Kumar J., J. Comput. Electron., 2016, \textbf{15}, No.~4,  1531--1540, \doi{10.1007/s10825-016-0911-5}.

\bibitem{Boyko2018}  Boyko I.V., Condens. Matter Phys., 2018, \textbf{21}, No.~4,  43701, \doi{10.5488/CMP.21.43701}.

\bibitem{Bayerl2019}  Bayerl D.,  Kioupakis E., Appl. Phys. Lett., 2019, \textbf{115}, No.~13,
        131101, \doi{10.1063/1.5111546}.
        
\bibitem{Staszczak2020} Staszczak G., Trzeciakowski W., Monroy E., Bercha A., Muziol G., Skierbiszewski C., Perlin P., Suski T., \\ Phys. Rev. B, 2020, \textbf{101}, No.~8,  085306, \doi{10.1103/PhysRevB.101.085306}.

\bibitem{Yan2003} Yan Z.W., Ban S.L., Liang X.X., Eur. Phys. J. B, 2003, \textbf{35}, No.~1,  41--47, \doi{10.1140/epjb/e2003-00254-8}.

\bibitem{Yamanaka2008} Yamanaka T., Alexson D., Stroscio M.A.,  Dutta M., Petroff P., Brown J.,  Speck J., \\ J. Appl. Phys., 2008,
\textbf{104}, No.~9, 093512 (10 pages), \doi{10.1063/1.3013885}.

\bibitem{Zhang2006} Zhang L., Surf. Rev. Lett., 2006,\textbf{13}, No.~1, 75--80, \doi{10.1142/S0218625X0600786X}.

\bibitem{Pokatilov2003} Pokatilov E.P., Nika D.L., Balandin A.A., Superlattices Microstruct., 2003, \textbf{33}, No.~3, 155--171,\\ \doi{10.1016/S0749-6036(03)00069-7}.

\bibitem{Pokatilov2004} Pokatilov E.P., Nika D.L., Balandin A.A. J. Appl. Phys., 2004, \textbf{95}, No.~10, 5626--5632, \doi{10.1063/1.1710705}.

\bibitem{Yang2013} Yang F.J., Ban S.L., Solid State Commun., 2013, \textbf{161}, No.~1, 5--8, \doi{10.1016/j.ssc.2013.02.015}.

\bibitem{Zhu2016} Zhu L., Luo H., J. Alloys Compd., 2016, \textbf{685}, No.~1, 619--625, \doi{10.1016/j.jallcom.2016.05.314}.

\bibitem{Wang2018} Wang J., Zhu L., Yin. W., Comput. Mater. Sci., 2018, \textbf{145}, No.~1, 14--23, \doi{10.1016/j.commatsci.2017.12.058}.


\bibitem{Boyko2020}  Boyko I.V.,  Petryk M.R., Fraissard J., Nano Express, 2020, \textbf{1}, No.~1,
        010009 (13 pages), \doi{10.1088/2632-959X/ab7cb2}.

\bibitem{Boyko2020-1} Boyko I., Petryk M., Fraissard J., Eur. Phys. J. B, 2020, \textbf{93}, No.~3,  57 (13 pages),\\  \doi{10.1140/epjb/e2020-100597-x}.

\bibitem{Tkach2014}  Tkach M.V., Seti Ju.O., Grynyshyn Y.B., Voitsekhivska O.M., Condens. Matter Phys., 2014, \textbf{17}, No.~2,  23704 (10 pages), \doi{10.5488/CMP.17.23704  }.

\bibitem{Gao2007}  Gao X., Botez D.,  Knezevic I., J. Appl. Phys., 2007, \textbf{101}, No.~6,  063101 (10 pages), \doi{10.1063/1.2711153}.

\bibitem{Piprek2007} Piprek J., Nitride Semiconductor Devices: Principles and Simulation, Wiley-VCH, Weinheim, 2007, \doi{10.1002/9783527610723}.

\bibitem{Hedin1971}  Hedin L., Lundqvist B.I., J. Phys. C, 1971, \textbf{4}, No.~14,  2064--2083.

\bibitem{Stroscio2001} Stroscio M.A., Dutta M., Phonons in Nanostructures, Cambridge University Press, Cambridge, 2001, \\\doi{10.1017/CBO9780511534898}.

\bibitem{Tkach2003} Tkach M.V., Quasiparticles in Nanoheterosystems, Ruta, Chernivtsi, 2003 (in Ukrainian).

\bibitem{Yan2014} Yan Q., Rinke P., Janotti A., Scheffler M. Van de Walle C.G., Phys. Rev. B, 2014, \textbf{90},
        No.~12,  125118 (11 pages), \doi{10.1103/PhysRevB.90.125118}.

\bibitem{Vurgaftman2001}   Vurgaftman I.,  Meyer J.R. J. Appl. Phys., 2001, \textbf{94}, No.~6,
        3675--3696, \doi{10.1063/1.3236533}.
\end{thebibliography}
\end{document}